\documentclass[sigconf]{acmart}
\usepackage{textcomp}
\usepackage{xcolor}
\usepackage{tikz}
\usepackage{amsmath}
\usepackage{array}
\usepackage{booktabs}
\usepackage{multirow}
\AtBeginDocument{%
  }

\setcopyright{acmlicensed}
\copyrightyear{2025}
\acmYear{2025}
\acmDOI{XXXXXXX.XXXXXXX}
\acmConference[XXX '26]{2026 XXX}{XXX XX--XX, 2026}{XXX XXX XXX, XXX}

\acmISBN{978-1-4503-XXXX-X/2018/06}

\begin{document}

\title{Generality Is Not Enough: Zero-Label Cross-System Log-Based Anomaly Detection via Knowledge-Level Collaboration}

\author{Xinlong Zhao}
\orcid{0009-0001-8230-1039}
\affiliation{
  \institution{Peking University}
  \city{Beijing}
  \country{China}
}
\email{xlzhao25@stu.pku.edu.cn}

\author{Tong Jia}
\authornotemark[1]
\orcid{0000-0002-5946-9829}
\affiliation{
  \institution{Peking University}
  \city{Beijing}
  \country{China}
}
\email{jia.tong@pku.edu.cn}

\author{Minghua He}
\orcid{0000-0003-4439-9810}
\affiliation{
  \institution{Peking University}
  \city{Beijing}
  \country{China}
}
\email{hemh2120@stu.pku.edu.cn}

\author{Ying Li}
\orcid{0000-0001-9667-2423}
\affiliation{
  \institution{Peking University}
  \city{Beijing}
  \country{China}
}
\email{li.ying@pku.edu.cn}

\renewcommand{\shortauthors}{Xinlong Zhao et al.}

\begin{abstract}
Log-based anomaly detection is crucial for ensuring the stability and reliability of software systems. However, the scarcity of labeled logs severely constrains the rapid deployment of existing methods to new systems. To address this issue, cross-system transfer has been recognized as an important research direction. State-of-the-art cross-system approaches can achieve strong performance when a small amount of labeled target logs is available, but several limitations remain: methods based on small models focus solely on transferring general knowledge and thus overlook the discrepancies and potential mismatches between general knowledge and the target system’s proprietary knowledge; methods based on LLMs can capture proprietary patterns but often rely on guidance from a few positive examples and are constrained by high inference cost and low inference efficiency. Existing LLM–small model collaboration strategies typically quantify sample complexity via the small model’s output uncertainty, assigning “simple logs” to the small model and “complex logs” to the LLM. However, in the cross-system setting without target labels, sample complexity cannot be measured in a supervised manner, and such routing mechanisms are not designed from the perspective of knowledge separation. Therefore, they are not directly applicable to zero-label cross-system log-based anomaly detection. To address these issues, we propose GeneralLog, a novel LLM–small model collaborative method for zero-label cross-system log-based anomaly detection. GeneralLog performs knowledge-level dynamic routing of unlabeled target logs, allowing the LLM to process “proprietary logs” while the small model handles “general logs,” thereby achieving cross-system generalization without requiring labeled target logs. Experiments on three public log datasets from different systems show that, under a fully zero-label setting, GeneralLog achieves an F1-score over 90\%, significantly outperforming existing state-of-the-art cross-system methods.
\end{abstract}





\maketitle

\begin{figure*}[h!]
\centering
\includegraphics[width=1.0\textwidth]{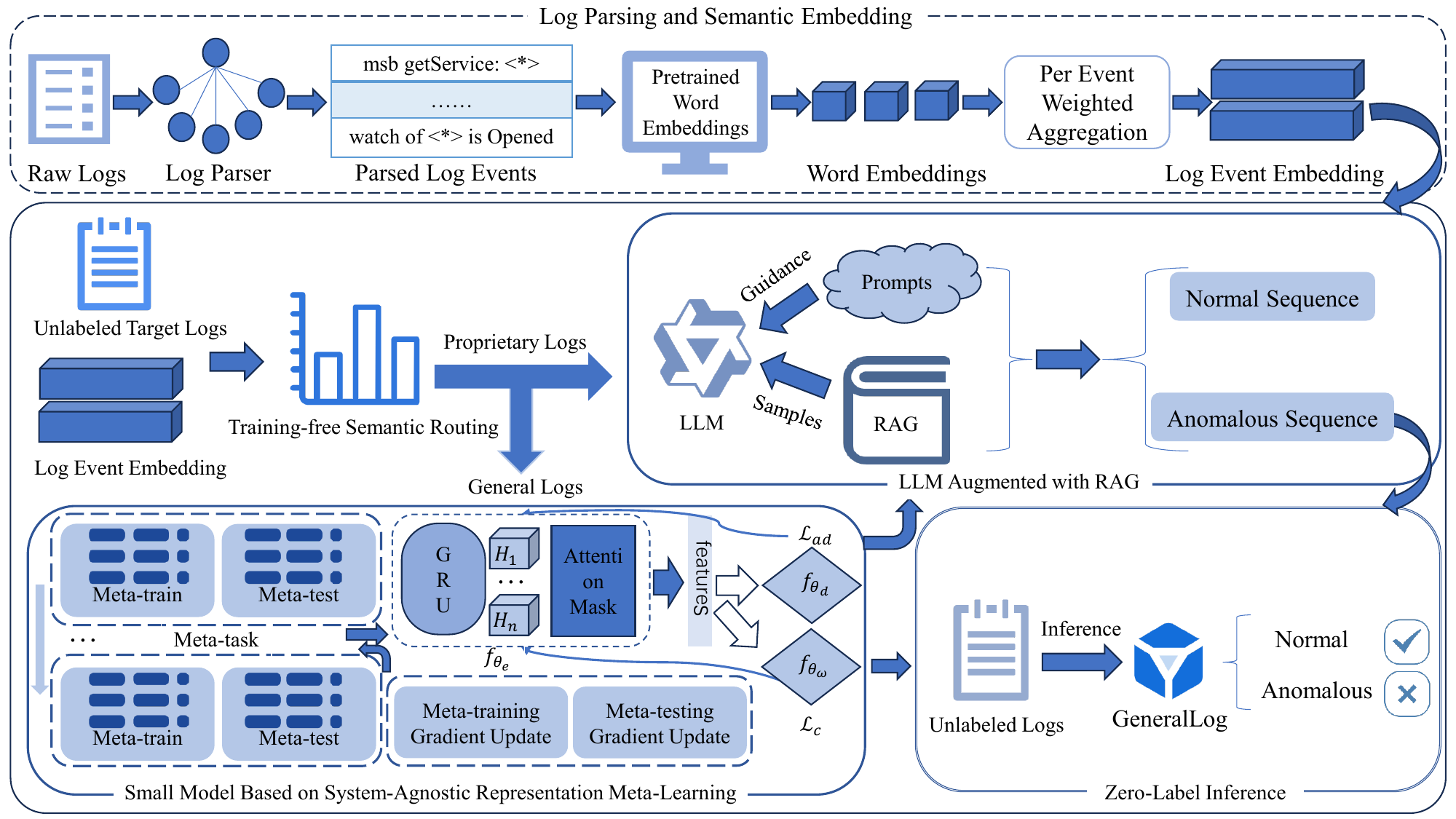}
\vspace{-0.5cm}
\caption{The proposed zero-label cross-system log-based anomaly detection pipeline for GeneralLog.}
\label{fig_2}
\end{figure*}

\section{Introduction}
As the scale and complexity of software systems continue to grow, the frequency of failures has shown an upward trend. Ensuring the reliability of systems has become one of the core challenges for their successful operation. System logs, which record key events and state changes, have become an essential source of information for anomaly detection~\cite{10.1609/aaai.v38i1.27764, Shajarian_2025, midlog, famos, yang2025enhancing, logcae, llmelog, afalog, aclog}. Log-based anomaly detection techniques have broad applications in improving system reliability. Existing anomaly detection models can mainly be divided into unsupervised and supervised models. Unsupervised models~\cite{10.1145/3377813.3381371, 9240683} use sequential neural networks to learn the occurrence probabilities of log events in normal event sequences, predicting subsequent log events and identifying events that deviate from the predictions as anomalies. However, due to the lack of explicit labeling of anomaly logs, the detection capability of these models is somewhat limited~\cite{9401970}. In contrast, supervised models~\cite{8854736, 10.1145/3338906.3338931} construct classification models to identify anomalous logs, typically demonstrating higher detection performance. However, their effectiveness relies heavily on a large number of labeled logs, which makes it highly challenging to apply supervised models to newly deployed software systems.

To address the above issues, researchers have proposed several cross‑system methods. Methods based on small model borrow general knowledge from mature systems via transfer learning~\cite{9251092, 10.1145/3459637.3482209} or meta-learning~\cite{10.1145/3597503.3639205, FreeLog}, thereby reducing reliance on large amounts of labeled logs. By contrast, LLM-based methods~\cite{10607047, eagerlog}, leveraging their strong intrinsic generalization capability, can achieve strong performance with only a few positive examples for guidance. However, existing studies have shown that transfer learning methods guarantee performance only under specific assumptions and may face substantial difficulties when the distribution discrepancy is significant~\cite{ijcai2022p496}. In contrast, Meta-learning involves external optimization, enabling model to handle broader meta-representations beyond just model parameters~\cite{9428530}. Existing research has shown that compared to transfer learning, meta-learning can achieve comparable generalization results with fewer data ~\cite{gu-etal-2018-meta}. However, whether based on transfer learning or meta‑learning, existing studies concentrate on extracting general knowledge from a global perspective and overlook the significant discrepancies at the level of proprietary knowledge between the source and target systems. Our observations of real-world logs reveal that entries exchanged between source and target systems fall into two distinct categories: semantically similar “general logs” and semantically dissimilar “proprietary logs”. In this context, general knowledge is derived from general logs, whereas proprietary knowledge is obtained from proprietary logs. Concretely, examples from two systems illustrate this dichotomy. HDFS logs record operations on the distributed file system, including file access, data replication, and node status, as well as associated warnings and error messages. These logs are primarily concerned with file system operations and are unrelated to hardware operations. BGL logs capture hardware status, task scheduling, and the execution of parallel computing jobs in a supercomputing environment. The first two templates exhibit the same operational pattern of timeouts and thus exemplify general logs: HDFS timeout template: "\texttt{<*>} PendingReplicationMonitor timed out block \(\texttt{<*>}\)" — reports a block replication timeout. BGL timeout template: "job \(\texttt{<*>}\) timed out" — reports a job timeout. Both templates share the surface phrase “timed out,” reflecting a shared failure mode across systems. By contrast, the following templates capture system-specific operations and therefore constitute proprietary logs. These entries have distinct semantics and structured placeholders that are unique to each system: HDFS proprietary template: "\texttt{<*>} Changing block file offset of block \texttt{<*>} from \texttt{<*>} to \texttt{<*>} meta file offset to \texttt{<*>}" — denotes an internal HDFS operation that modifies a block’s file offset; the template contains placeholders for block ID, original offset, new offset, and the meta file offset. BGL proprietary template: "Bad cable going into LinkCard \texttt{<*>} Jtag \texttt{<*>} Port \texttt{<*>} - \texttt{<*>} bad wires" — reports a hardware-level cable fault in the supercomputing environment, with placeholders for LinkCard identifier, JTAG identifier, port identifier, and cable status. These examples demonstrate a clear disparity and mismatch between cross-system general knowledge and a target system’s proprietary knowledge. While general logs enable transfer of shared operational patterns, proprietary logs encode system-specific semantics and formats that are not captured by global generalizations. This divergence impedes a model’s ability to learn the target system’s proprietary knowledge, thereby constraining cross-system generalization and overall performance. For LLM-based methods, although capable of capturing proprietary patterns, they typically rely on guidance from a small number of positive examples and are constrained by high inference costs and low inference efficiency. These issues motivate us to seek an LLM–small model collaboration strategy to address cross-system log-based anomaly detection. However, existing LLM–small model collaboration methods typically rely on the small model’s output uncertainty to quantify the complexity of log entries, assigning “simple logs” to the small model and “complex logs” to the LLM. Such uncertainty-based routing mechanisms are difficult to apply in cross-system scenarios with unlabeled target logs: log complexity cannot be reliably quantified in unsupervised settings, and current routing are not designed from a knowledge-centric perspective, these schemes are unsuitable for zero-label cross-system log-based anomaly detection. This observation motivated our research.

To address the above challenges, we conceptualize cross‑system logs at the knowledge level by dividing them into “general logs” and “proprietary logs,” and to design specialized processing strategies based on their distinct characteristics. This breaks away from existing approaches that focus solely on general log patterns, ignore proprietary logs of the target system. Specifically, we propose GeneralLog, a novel LLM–small model collaborative method for zero-label cross-system log-based anomaly detection. To enable accurate classification and routing of logs, we design training-free router based on log semantic similarity that dynamically partitions the unlabeled target logs into “general logs” and “proprietary logs,” thereby providing fine‑grained inputs for subsequent processing. For general logs, we adopts a small model based on system‑agnostic representation meta‑learning for direct training and inference, inheriting general patterns between the source and target systems. For proprietary logs, we leverage an LLM augmented with a retrieval-augmented generation (RAG) knowledge base, which is constructed from the target system’s general logs and the labels predicted by the small model, to directly perform reasoning and identify system-specific anomaly patterns. We evaluate the performance of GeneralLog on three public log datasets from different systems (HDFS, BGL and OpenStack). Results show that under zero-label conditions, GeneralLog achieves over 90\% F1-score, significantly outperforming state‑of‑the‑art cross‑system log-based anomaly detection methods.

\begin{table*}[h!]
\centering
\setlength{\belowcaptionskip}{0.3cm}
\caption{Zero-label generalization experiments across different domains.}
\vspace{-0.2cm}
\label{tab_all}
\resizebox{\textwidth}{!}{%
\begin{tabular}{@{}lcccccccccccc@{}}
\toprule
\multicolumn{1}{c}{\multirow{2}{*}{\textbf{Method}}} & \multicolumn{3}{c}{\textbf{HDFS to BGL}} & \multicolumn{3}{c}{\textbf{BGL to HDFS}} & \multicolumn{3}{c}{\textbf{OpenStack to HDFS}} & \multicolumn{3}{c}{\textbf{OpenStack to BGL}} \\
\cmidrule(lr){2-4} \cmidrule(lr){5-7} \cmidrule(lr){8-10} \cmidrule(lr){11-13}
 & \textbf{Precision} & \textbf{Recall} & \textbf{F1-score} & \textbf{Precision} & \textbf{Recall} & \textbf{F1-score} & \textbf{Precision} & \textbf{Recall} & \textbf{F1-score} & \textbf{Precision} & \textbf{Recall} & \textbf{F1-score} \\

\midrule
PLELog (a1)  & 82.10 & 67.42 & 74.04 & 65.86 & 71.11 & 68.38 & 65.86 & 71.11 & 68.38 & 82.10 & 67.42 & 74.04 \\
LogRobust (a2)  & 94.60 & 72.95 & 82.38 & 100.00 & 62.30 & 76.77 & 100.00 & 62.30 & 76.77 & 94.60 & 72.95 & 82.38 \\ 
\hline
PLELog (b1)  & 94.88 & 89.62 & 92.18 & 96.30& 83.81& 89.62 & 96.30& 83.81& 89.62 & 94.88 & 89.62 & 92.18 \\
LogRobust (b2)  & 97.52 & 91.27 & 94.29 & 82.54& 99.20& 90.11 & 82.54& 99.20& 90.11 & 97.52 & 91.27 & \underline{94.29} \\ 
\hline
LogTAD (c1)  & 78.01 & 68.51 & 72.95 & 78.80 & 71.22 & 74.82 & 71.89 & 65.51 & 68.55 & 70.72 & 65.51 & 68.02 \\
LogTransfer (c2)  & 74.42 & 76.73 & 75.56 & 100.00 & 43.30 & 60.43 & 73.87 & 62.30 & 67.59 & 68.43 & 71.32 & 69.85 \\ 
LogDLR (c3)  & 79.25 & 72.56 & 75.76 & 77.78 & 70.05 & 73.71 & 73.65 & 69.80 & 71.67 & 69.58 & 64.33 & 66.85 \\ 
\hline
MetaLog (d1)  & 96.89 & 89.28 & 92.93 & 89.29 & 74.98 & 81.51 & 96.67 & 62.42 & 75.86 & 99.83 & 70.09 & 82.36 \\
MetaLog (d2)  & 98.75 & 20.05 & 33.33 & 72.29 & 12.10 & 20.73 & 100.00 & 18.35 & 31.01 & 100.00 & 1.20 & 2.37 \\  
\hline
DeepLog (e1)  & 66.13 & 48.79 & 56.16 & 53.96 & 34.07 & 41.77 & 53.96 & 34.07 & 41.77 & 66.13 & 48.79 & 56.16 \\ 
\hline
PLELog (f1)  & 38.80 & 99.87 & 55.89 & 1.69 & 92.85 & 3.32 & 4.33 & 53.47 & 8.01 & 54.65 & 43.04 & 48.16 \\
LogRobust (f2)  & 39.08 & 93.67 & 55.15 & 2.25 & 62.12 & 4.35 & 0.63 & 60.81 & 1.25 & 34.34 & 57.39 & 42.97 \\
MetaLog (f3)  & 29.43 & 0.45 & 0.89 & 2.90 & 80.92 & 5.61 & 2.90 & 80.92 & 5.61 & 29.43 & 0.45 & 0.89 \\
FreeLog (f4)  & 81.12 & 77.10 & 79.06 & 79.34 & 76.11 & 77.69 & 71.34 & 79.09 & 75.02 & 73.59 & 80.33 & 76.81 \\ 
\hline
RAGLog (g1)  & 81.25 & 98.50 & 89.05 & 85.33 & 97.65 & 91.08 & 85.33 & 97.65 & 91.08 & 81.25 & 98.50 & 89.05 \\
RAGLog (g2)  & 45.76 & 100.00 & 62.79 & 49.50 & 100.00 & 66.22 & 49.50 & 100.00 & 66.22 & 45.76 & 100.00 & 62.79 \\
\hline
Ours GeneralLog & 94.35 & 96.64 & \textbf{95.48} & 93.28 & 96.53 & \textbf{94.88} & 90.18 & 93.85 & \textbf{91.98} & 91.74 & 92.88 & \textbf{92.31} \\
\bottomrule
\end{tabular}
}
\end{table*}

\section{Method}
\label{sec3}
GeneralLog comprises three core components: Training-free Semantic Routing, Small Model Based on System-Agnostic Representation Meta-Learning, and LLM augmented with RAG. Through their synergistic collaboration, GeneralLog is able to concurrently capture both general and proprietary knowledge in a zero‑label cold‑start scenario. The complete workflow is illustrated in Figure \ref{fig_2}.

GeneralLog begins by employing the classic log parsing method Drain~\cite{8029742} to process unstructured raw logs from various software systems and extract log events. To account for the cross-system nature, we adopt the semantic embedding approach inspired by MetaLog~\cite{10.1145/3597503.3639205}, which ensures consistency in event representations by constructing semantic embedding vectors for log events within a shared global space. After obtaining semantic embeddings for each log event, GeneralLog then performs semantic routing to partition the unlabeled target log sequences into “general logs” and “proprietary logs.” The procedure is as follows: Let a target log sequence be $x_k = \{l_1, l_2, \dots, l_n\}$, where each log entry $l_i$ has an event embedding vector ${v}_i \in \mathbb{R}^d$. Denote the set of all event embeddings of source system by ${V}^{source} = \{u_1, u_2, \dots, u_m\}, u_j \in \mathbb{R}^d$. Compute the cosine similarity between ${v}_i$ and each ${u}_j$, and take the maximum: $sim_i = \max_{1 \le j \le m} \mathrm{cosine}({v}_i, {u}_j) = \max_{j} \frac{{v}_i \cdot {u}_j}{\|{v}_i\|\;\|{u}_j\|}$. To assess the overall similarity of sequence $x_k$ to the source domain, aggregate by taking the minimum event‑level score: ${x_k}_{\mathrm{sim}} = \min_{1 \le i \le n} sim_i$. This highlights the most representative event in the sequence and, since $|{V}^{source}|$ is only on the order of tens or hundreds, keeps computation efficient. Given a threshold $\tau \in [0,1]$, assign sequence $x_k$ as: \(x_k \in \text{General Logs if } {x_k}_{\mathrm{sim}} \ge \tau\) , otherwise Proprietary Logs. Through this semantic routing process, GeneralLog accurately separates sequences that share patterns with the source system from those exhibiting system‑specific behaviors, providing fine‑grained inputs for the subsequent modules.

In zero-label cross-system scenario, FreeLog ~\cite{FreeLog} achieves effective extraction and generalization of general knowledge through a system-agnostic representation meta-learning method. Drawing inspiration from FreeLog’s small model design, for general logs, GeneralLog adopts a small model based on system‑agnostic representation meta‑learning for direct training and inference. The small model of GeneralLog consists of two key stages: adversarial unsupervised domain adaptation and meta-learning. Specifically, we use \( X_S \) and \( X_T \) to represent the logs sampled from the source domain \( D_S \) and the target domain \( D_T \), while \( Y_S \) denotes the label matrix for \( X_S \). We first construct a cross-system meta-task \( MT_i = \{ M_i^{sup}, M_i^{que} \}\), where \( M_i^{sup} = \{ X_{S_i}^{sup}, X_{T_i}^{sup}, Y_{S_i}^{sup} \} \) is used for meta-training, \( M_i^{que} = \{ X_{S_i}^{que}, X_{T_i}^{que}, Y_{S_i}^{que} \} \) is used for meta-testing, and \( X^{sup} \) and \( X^{que} \) are referred to as the support and query set. The small model of GeneralLog consists of three main modules: the feature extractor \( f_{\theta_e} \), the anomaly classifier \( f_{\theta_\omega} \) and the domain classifier \( f_{\theta_d} \). Like Metalog~\cite{10.1145/3597503.3639205} and FreeLog~\cite{FreeLog}, \( f_{\theta_e} \) consists of two modules: the Gated Recurrent Unit (GRU) and the attention mask layer. Given a sequence of log event embeddings, the GRU module maintains a hidden state at each time step, enabling the network to retain long-term information from the input log event sequence. For each time step, the attention module takes the hidden states as input and utilizes adaptive self-attention to fuse the information. In each meta-task, we train the domain classifier \( f_{\theta_d} \) to maximize the distinction between features from the source and target domains. The optimization problem is as follows: $\max_{\theta_d} \sum_{MT_i} {L}_{ad}^{MT_i}(M_i^{sup}; f_{\theta_d})$, where the adversarial loss function is the binary cross-entropy with logits loss. Then, we train the anomaly classifier \( f_{\theta_\omega} \) to learn discriminative features for classifying normal and anomalous logs. The optimization problem is as follows: $\min_{\theta_\omega} \sum_{MT_i} {L}_c^{MT_i}(M_i^{sup}; f_{\theta_\omega})$,  where the classification loss function is the binary cross-entropy loss. During the meta-training phase, the learner's parameters can be updated through one or more gradient descent steps: $\theta_e^i = \theta_e - \delta \nabla_{\theta_e} {L}_{MT_i}(M_i^{sup}; f_{\theta_e})$, where \( \delta \) is the learning rate and the objective function ${L}_{MT_i}(f_{\theta_e}) = \gamma {L}_c^{MT_i}(X_{S_i}^{sup}, Y_{S_i}^{sup}; f_{\theta_\omega}) - \beta {L}_{ad}^{MT_i}(X_{S_i}^{sup}, X_{T_i}^{sup}; f_{\theta_d}).$ The hyperparameters \( \beta \) and \( \gamma \) control the trade-off between adaptation and classification performance. After learning the adaptation parameters \( \theta_e^i \) for each task, we proceed to meta-optimize \( f_{\theta_e} \) to improve the performance of \( \theta_e^i \) on the query set. The meta-objective function can be expressed as: $\min_{\theta_e} \sum_{MT_i} {L}_{MT_i}(M_i^{que}; f_{\theta_e^i})$. We perform meta-optimization via gradient descent as follows: $\theta_e \gets \theta_e - \alpha \nabla_{\theta_e} \sum_{MT_i} {L}_{MT_i}(M_i^{que}; f_{\theta_e^i}),$ where \( \alpha \) is the meta-step size. In summary, in each meta-task we learn general knowledge between labeled source logs and unlabeled target logs via adversarial unsupervised domain adaptation, and train a classifier on the source data to acquire discriminative information. This training strategy balances domain alignment and discriminative capability, enabling the model to maintain strong generalization and performance.

For proprietary logs, we employ an LLM (Qwen3) augmented with a retrieval-augmented generation (RAG) knowledge base built from the target system’s general logs and the small model’s highly accurate labels to perform reasoning and identify system-specific anomaly patterns ~\cite{10607047}. The small model produces precise labels for the general logs it excels at, and these in-domain positive exemplars provide strong guidance for interpreting proprietary behaviors. By combining the LLM’s powerful generalization with RAG-based retrieval of such labeled, context-rich examples, GeneralLog can robustly identify anomaly patterns unique to the target system while reducing costs.

\section{Experiments}
\label{sec4}
\textit{Datasets.} We conducted systematic experiments on three publicly log datasets: HDFS~\cite{10.1145/1629575.1629587}, BGL~\cite{4273008} and OpenStack~\cite{10.1145/3133956.3134015}. In zero-label setting, we selected four cross-system dataset combinations (HDFS to BGL, BGL to HDFS, OpenStack to HDFS and OpenStack to BGL) to validate our method. For the four experimental setups, we followed the code provided in~\cite{9401970} and used Drain~\cite{8029742} to parse the log events and organize the log sequences. This preprocessing ensured that the structures of all datasets were consistent with previous research methods, thus enabling a fair comparison.

\textit{Baselines.} Table \hyperref[tab_all]{1} evaluates four categories of baselines: Supervised baselines (blocks a–b): PLELog and LogRobust trained on varying fractions of target logs to assess performance under scarce (1\% anomaly) versus fully labeled settings. Cross‑system baselines (blocks c–d): transfer learning and meta‑learning methods, each tested with partially labeled source and target logs as well as in zero anomaly label variants to gauge generalization across datasets. Zero‑label baselines (blocks e–f): DeepLog trained only on normal target logs, alongside zero‑label generalization of other method trained exclusively on source logs and evaluated directly on the unlabeled target. LLM baseline (block g): RAGLog evaluated with 10\% of target logs in the RAG knowledge base and, alternatively, without RAG integration of labeled target logs.

% \vspace{-0.3cm}
\begin{figure}[h!]
\centering
\includegraphics[width=1.0\columnwidth]{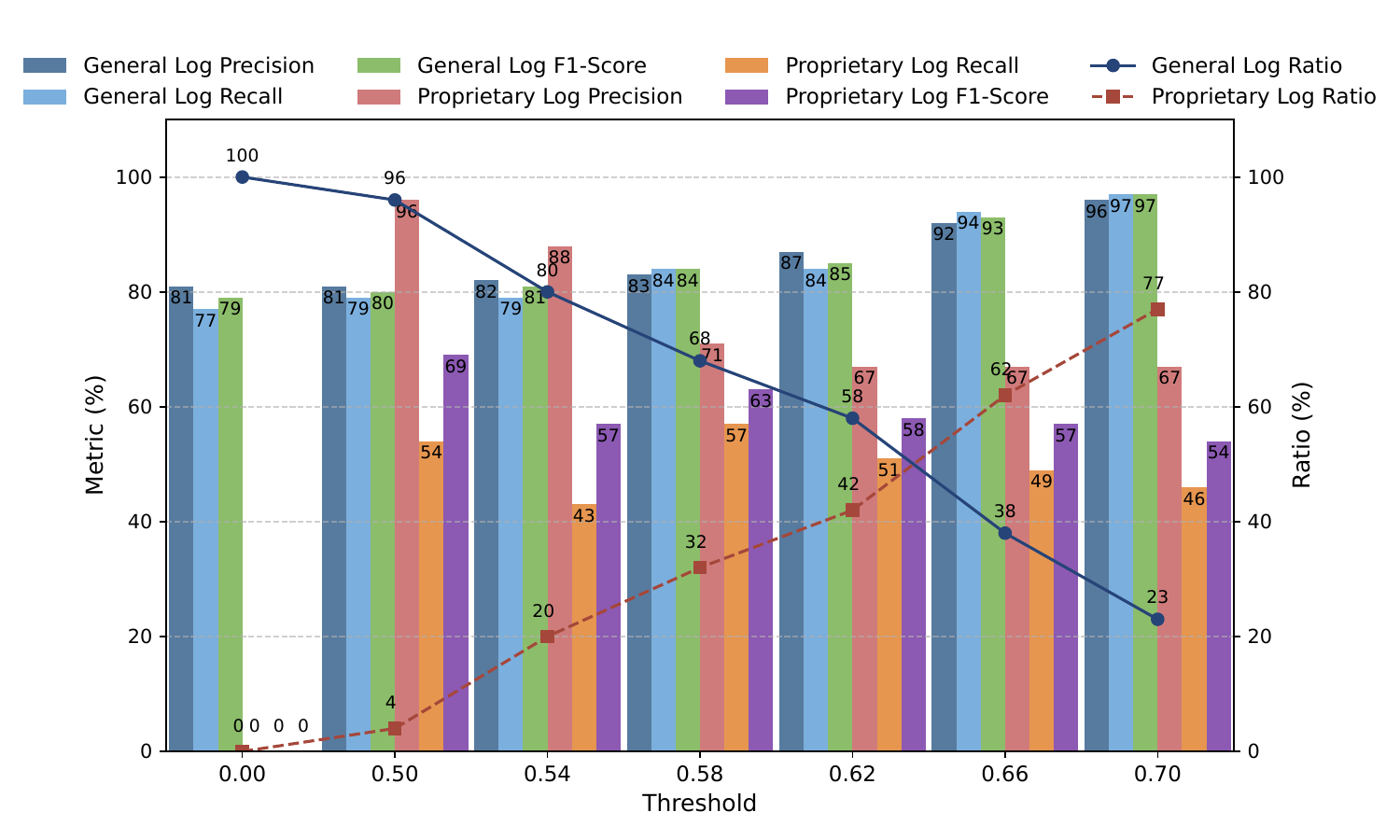}
\vspace{-0.75cm}
\caption{Routing Analysis.}
\label{fig_3}
\end{figure}
% \vspace{-0.45cm}

\textit{Experimental Analysis} Table \hyperref[tab_all]{1} shows the comparative results of GeneralLog and various methods under four settings. As shown in the table, existing methods cannot fully address the zero‑label cold-start problem, whereas GeneralLog, through its design combining semantic routing and LLM-small model collaboration, achieves an F1‑score that significantly outperforms all baselines. In addition, we conducted ablation experiments in the HDFS to BGL transfer setting to evaluate the effectiveness of the training-free semantic routing module. By adjusting the similarity threshold of the router, we analyzed the log partitioning results and the small model’s performance on all logs, general logs, and proprietary logs. As shown in Figure \ref{fig_3}, as the threshold increases, its accuracy on general logs remains high, while its performance on proprietary logs drops significantly. Sensitivity analysis further shows that higher similarity thresholds can filter out purer general logs and improve the F1-score on general logs, indicating that the router can effectively distinguish cross-system shared patterns from system-specific behaviors.

\section{Future Plans}
\label{sec6}
Building on the promising results of GeneralLog, we outline several directions for future research to further enhance cross-system zero-label log-based anomaly detection. First, we aim to explore novel semantic routing mechanisms to more accurately distinguish between general and proprietary logs, thereby improving the precision of log partitioning and boosting anomaly detection performance. Second, future work will focus on integrating proprietary knowledge directly into the small model, enabling full utilization of the LLM’s knowledge while significantly reducing inference costs and improving efficiency. Finally, we plan to extend the method to incorporate proprietary knowledge from diverse data modalities, such as source code, system documentation, and configuration files, to enhance the detection of system-specific anomalies.

% \begin{acks}
% \end{acks}

\bibliographystyle{ACM-Reference-Format}
\bibliography{reference}


\end{document}